# Resistive MSGC with double layered electrodes

R. Oliveira, S. Franchino, V. Peskov

CERN

**Abstract**

The first successful attempts to optimize the electric field in Resistive Microstrip Gas Chamber (RMSGC) using additional "field shaping" strips located inside the detector substrate are described.

## I. Introduction

Recently a new approach was applied to micropattern detectors design and manufacturing: a multilayer printed circuit technology widely used nowadays in microelectronics (see [1-3]). First it was implemented in micropattern detectors with resistive electrodes. These modified detectors contain additional strips located ether under the resistive electrodes [2] or inside the detector substrate [3] and were used to detect signals produced by avalanches in the amplification gap. Such a double layer structure ensures protection of the front end electronics, connected to the inner strips, preventing it from being damaged in the case of occasional sparks.

In the electrostatic case a double layer electrode system offers an extra possibility of optimizing the electric field in the avalanche region. Of course, in the presence of avalanche ions attached to the dielectric substrate surface the situation will be dynamic: ions deposited on the surface can temporaly change the local electric field. However after some time:

$$\tau \sim \varepsilon_0 \varepsilon \rho \quad (1),$$

(where $\varepsilon_0$ and $\varepsilon$ are the permittivity of the vacuum and the intermediate layer between the active electrode and the inner strips respectively and $\rho$ is the latter resistivity). When the charge dissipates, the field line map will return to the initial one. Hence one can expect that the inner electrodes can efficiently influence the field only at a counting rate below some critical value $1/\tau$.

The aim of the work was to investigate the feasibility of electric field tuning in a double layered micropattern detector by applying voltages to its inner strips. This can be useful in some cases, for example when the surface streamers limit the maximum achievable gas gains [4,5]. In the latter case an effective measure is to minimize the field parallel to the surface. For this reason MSGCs were chosen for thesestudies and both simulations and measurements were carried out.

## II. Simulation

We have simulated electric fields in the device using COMSOL Multiphysics with the electrostatic package.

The detector geometry used for the simulation is shown in Fig.1. It is a double layer MSCG having anode and cathode strip widths of 100 μm and 680 μm respectively and a pitch of 1.5 mm. At 100 µm below these, inside the substrate, an additional array of strips is located oriented parallel to the top electrodes.

For comparison, a simulation with the same surface geometry, but without embedded electrodes was also performed; this is like a standard MSGC having, however a larger anode width and pitch.

The simulations were done for various voltages applied to the inner electrodes located under the cathode strips. As examples, results of simulations for two voltage settings are shown below:

1) 1100 V applied to the anode strip with all other electrodes  grounded (Fig. 2a)

2) in addition, -200 V are applied to the electrodes located under the cathode strip (Fig. 2b).

In both cases the drift field was 1.5 kV/cm.

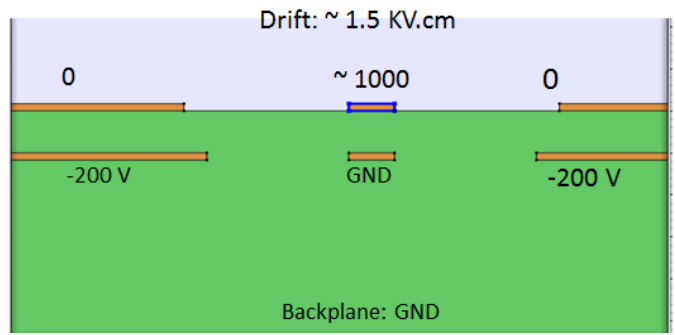

Figure 1. MSGC geometry used for electrostatic simulations

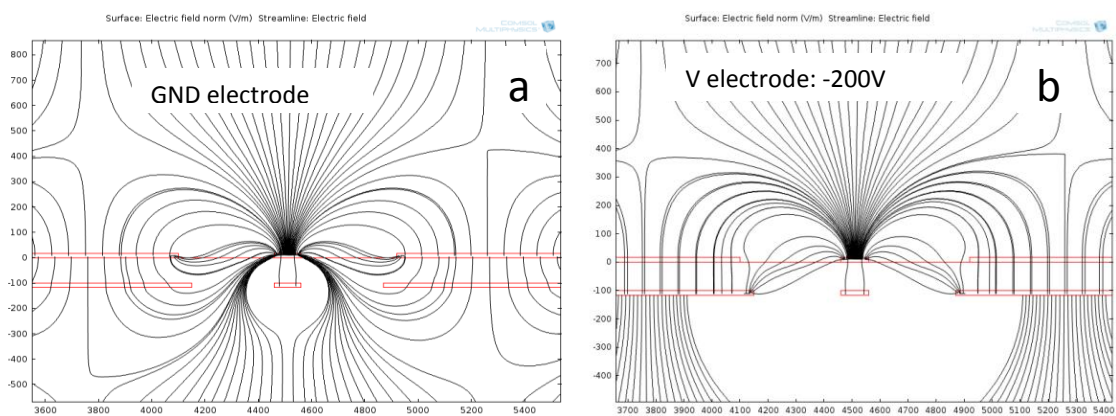

Figure 2. Simulated field lines in the device: a) the case when 1100 V is applied to the anode strip, whereas all other electrodes are grounded. b)the  case when -200V is applied to the inner electrodes located under the cathode strips.

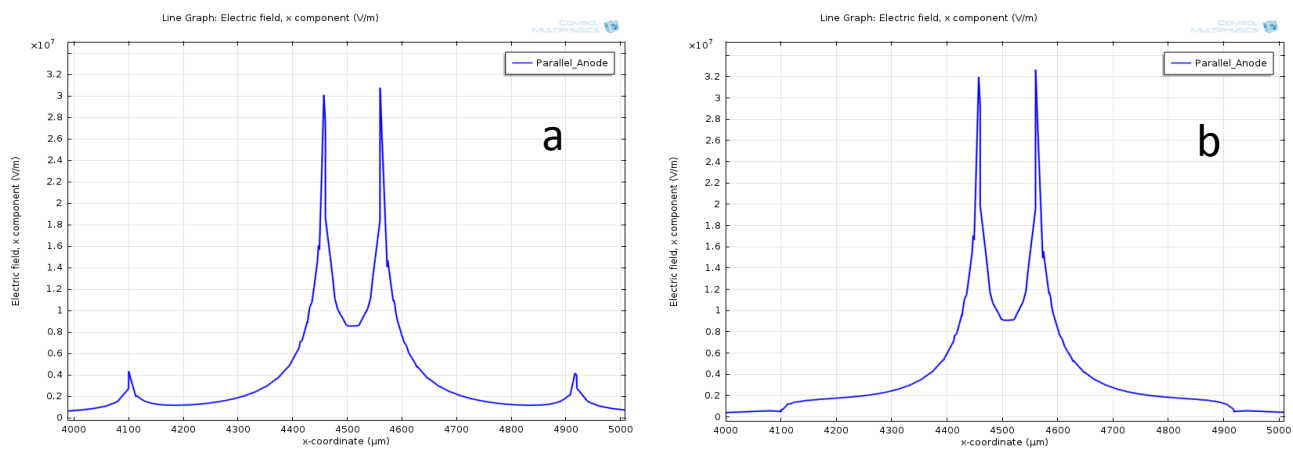

Figure 3.Electric field scan in a direction perpendicular to thestrips: a) grounded internal electrode; b) -200V applied to the inner cathode strips. Two large peaks in the center of the figure correspond to the electric field in the anode region. Two other peaks are visible at the cathode edges in the case in which internal strips are grounded.

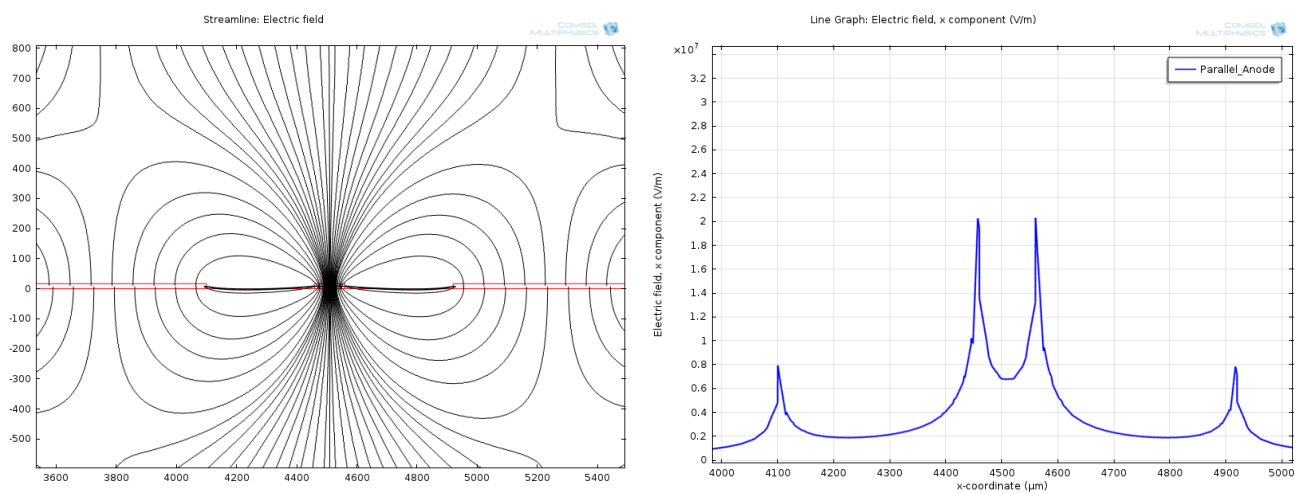

Figure 4. Simulated field lines (left) and horizontal electric field (right) for the device without embedded strips. High electric field at the edges of cathodes is clearly visible.

As can be seen from Fig. 3a, when the inner electrodes were grounded, the electric field on the MSGC surface has a high value not only in the anode region, but also near the cathode edges.

This effect is more pronounced in the case of the standard MSGC, without embedded electrodes, as can be seen from Fig. 4. These results are similar to earlier simulations made by various authors (see for example [5]).

Dramatic changes happened when -200V were applied to the inner electrodes (Fig 3b): in this case the electric field near the anode strips increased, whereas the peaks near the cathode strips almost disappeared. Such field modifications may lead to an increase of the maximum achievable gain of the MSGC. Indeed, as was shown in [4,5] ,usually surface streamers limit the maximum achievable gain in MSGCs. The favorable factors for their formation are:

1) High value of electric field near the edges, where some streamers originate,

and

2) High enough electric field component along the substrate surface necessary to support the streamer propagation. More electric field lines near the anode strip resembling "radial" type geometry (as in a single wire detector), means less probability for the streamer to develop and propagate [5].

Of course, as was mentioned earlier, these calculations are purely electrostatic and in a real situation there could be a significant contribution from avalanche ions attached on the substrate surface and modifying the electric field

In any case, in practice, one cannot easily verify the calculated field line maps. However, one can try to find some indirect indications. For example, one can measure the maximum achievable gain as a function of the voltages applied to the inner strips and check if it really increases. Moreover, if the anode voltage can be increased before breakdown appears (with respect to the case with grounded strips), one can speculate that the electric field near the cathode edges probably decreased.

This was the aim of the experimental part of this work

## III. Manufacturing of RMSGC

The RMSGC was manufactured at the printed circuits workshop at CERN as a standard multilayer Printed Circuit Board (PCB) using the photolithography technique.

The manufacturing process started with a 1.6 mm thick fiber glass plate (FR4, EMC 370) with 35 μm of copper on both sides. On the top of it, readout strips (100 μm wide) and corrective/"inner" electrode strips (780 μm wide) were chemically etched (Fig.5 a).On the bottom of it, the entire layer of Cu was preserved, in order to act as the grounded backplane of the final detector.

Then, a 100 μm thick dielectric (glass fiber and epoxy glue) and 17 μm of Cu were pressed on top of the first PCB layer (Figure 5 b).

Resistive cathodes have been created on this copper layer. For this, parallel grooves, 680 μm wide, were etched on the top Cu layer (Fig 5 c), and then filled with resistive paste with surface resistivity of 1 MΩ/□ (ESL RS 12116), as shown in Figure 5 d.

The anode strips were then created in the remaining copper on the top surface of the PCB, chemically etching the unwanted metal (figure 5e).

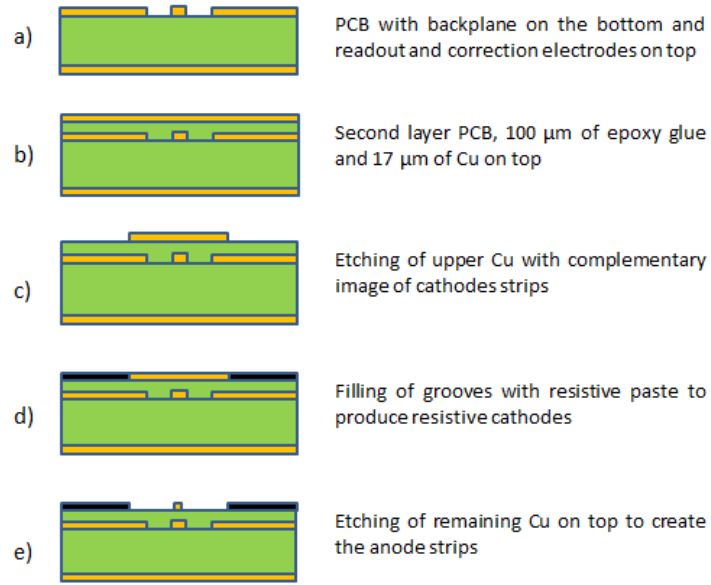

Figure 5. Manufacturing steps of RMSGC.

Two detectors were produced: one with 30 μm wide anode strips and one with 100 μm ones. The strip's pitch for both detectors was 1.5 mm and the total active area was 3*3 cm$^2$. Some photographs of the final detector can be seen in Figures 6 and 7.

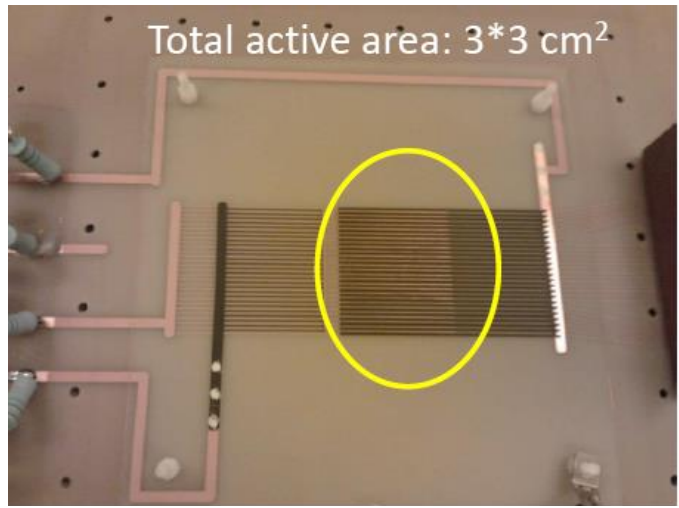

Figure 6. Photograph of the final RMSGC detector, the active area is emphasized by the yellow circle.

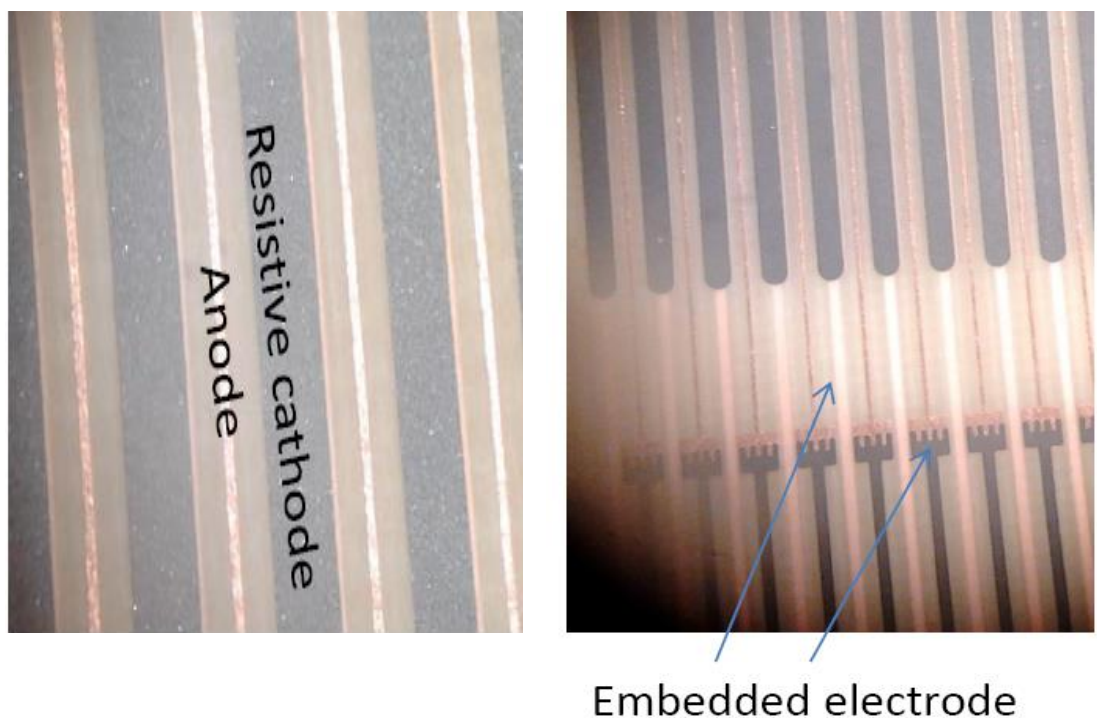

Figure 7. Magnified photographs of the RMSGC. One can see resistive cathodes (in black), Cu anodes and the embedded electrode strips below the resistive ones.

## IV. Experimental setup

The experimental setup used in these studies is shown schematically in Fig.8. It consists of a gas chamber housing one of the RMSGCs described above, a gas system and electronics. 10 mm above the MSGC a drift electrode was placed. The gas chamber was flushed with a mixture of Ar with 30% of $CO_2$ at a total pressure of 1 atm. It had a window transparent to 6 keV photons and correspondingly a $^{55}$Fe source was used to produce primary electrons in the drift region, although some studies were performed with an $^{241}$Am source placed inside the gas chamber.

To minimize the number of electronic channels all anode and cathode strips were connected together and to the common anode and cathode electrode respectively (see Fig.6). In most measurements the cathode electrode was kept grounded and the high voltage was applied to the anode electrode via the charge sensitive preamplifier CAEN A422A. The signals from the preamplifier were further amplified and shaped with a research amplifier from ORTEC and were then sent to a pulse-height analyzer MCA of Amptek which recorded the spectrum of the radioactive source and also determined its counting rate which was seen to be was comparable to the one measured with a scaler. If necessary, signals from the inner anode strips could be detected as well. In measurements of the gas gain a picoammeter was used. From the measured anode current and the counting rate, the gas gain was determined assuming that each 6 keV photon produced about 200 primary electrons.

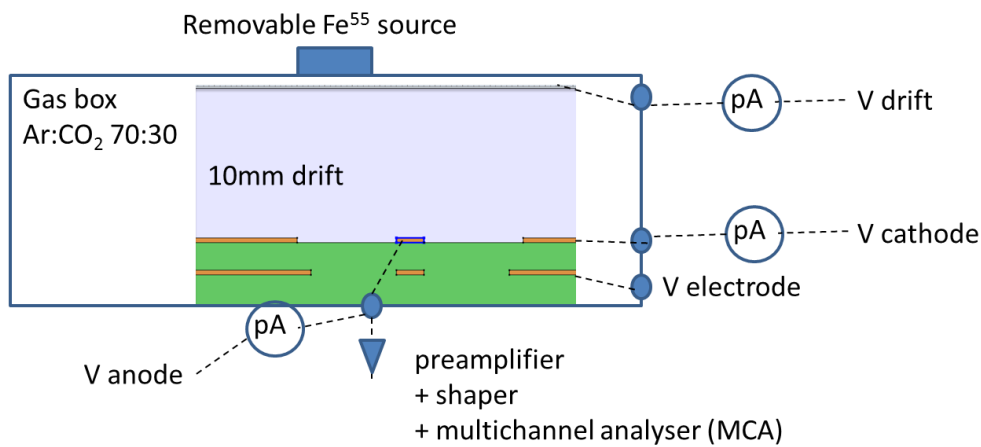

Figure 8. Schematic drawing of the experimental setup

## V. Preliminary results

Fig.9 shows results of RMSGC gain measurements performed at various voltages applied to the inner cathode strips $V_{el}$ with a detector with 30 μm strips.

It is clearly seen that the gas gain increases with $V_{el}$ reaching values close to $10^5$ which is ten times higher than those achieved earlier with an ordinary RMSGC [6]. The influence of $V_{el}$ on the gas gain, for a given anode voltage (950V) can be seen more easily in Fig.10.

From the data presented in Fig .11 one can see the effect of the voltage applied in the inner electrode on the maximum achiavable anode voltage before a spark. A spark was defined as a trip of the power supply, where the maximum current was set to 50 nA. Each point in the plot was averaged over 6 sparks for each voltage configuration. The effect of the inner electrode is clearly visible; it allows a voltage on the anode that is 100V higher to be reached. The optimal value seems to be Vel=-200, which is compatible with the simulation (Fig. 3b). In fact,with this voltage configuration, in fact, the simulation shows almost no electric field peaks at the edges of the cathode, predicting that a higher voltage can be applied between the anode and the cathode before streamers appear.

Thanks to the resistive protection there was no visible damage to the device after sparking.

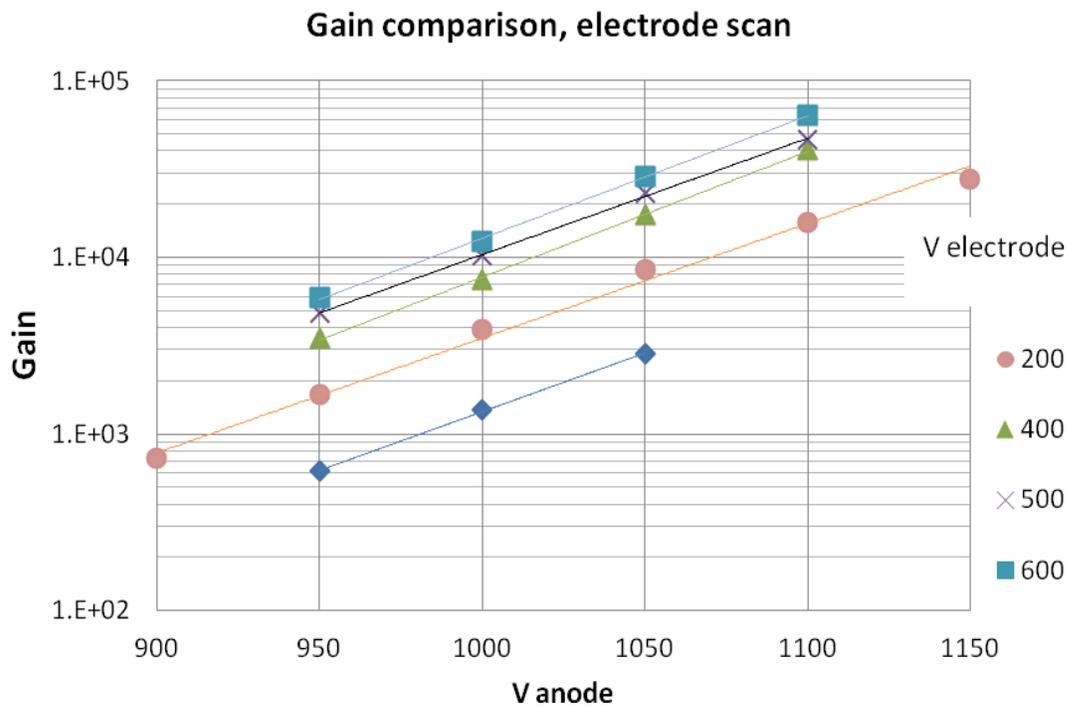

Figure 9. Gas gains of RMSGC vs. the anode voltage measured at different negative $V_{el}$. Detector with 30 µm strips.

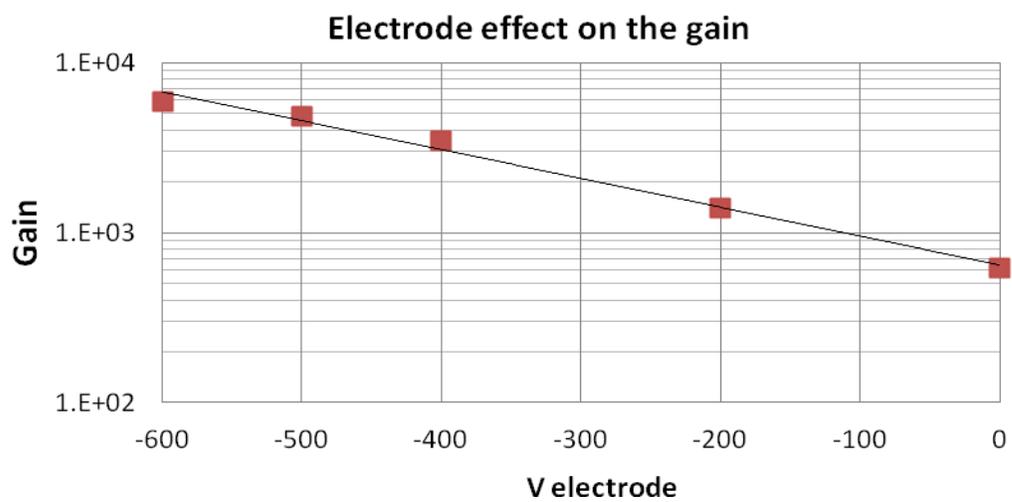

Figure 10. Gain dependence on $V_{el}$ for V anode of 950V in the detector with 30 µm strips.

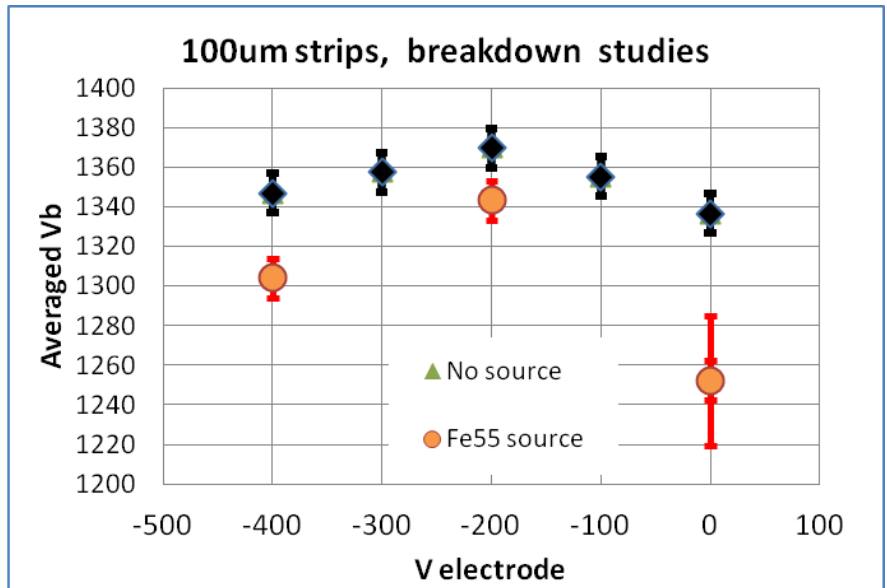

Figure 11. Effects on maximum achiavable anode voltage for different $V_{el}$ with the detector with100 μm strips.

The typical pulse-height spectrum of the $^{55}$Fe measured with RMSGC is depicted in Fig.12

 The energy resolution  was between 26 and 28%, depending on the electrode voltage, which is comparable to earlier results  obtained with  a similar RMSGC [6].

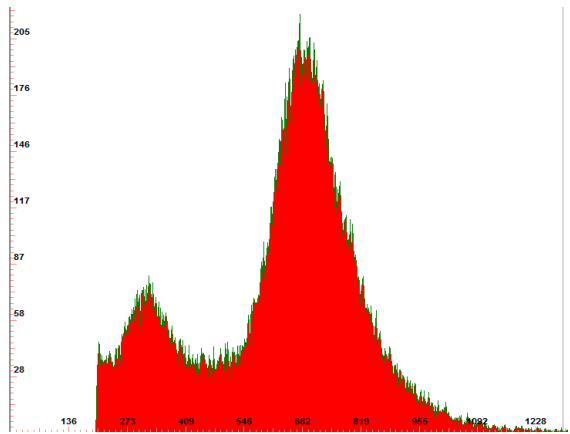

Figure 12. Pulse height spetrum of $^{55}$Fe  measured with RMSGC at $V_{el=}$0. Similar  spectra were recorded at various $V_{el}$.

## VI. Discussion and future plans

Preliminary results indicate that the effect of inner strips on the maximum achievable gain was clearly seen at least at the low counting rates used in our work. This proves that the surface streamers were efficiently suppressed allowing higher gains to be achieved and allowing measurements to be carried out in a region far away from were one start to see instabilities.

It must be noted that in these preliminary tests the intermediate layer between the detector active electrodes and the inner strips was not optimized: at present it is made of fiber glass FR4 which has a too high resistivity to prevent the detector from operating consistently at high counting rates.  Our previous experience with thick GEMs (TGEMs)  made of fiber glass show that two effects may contribute towards the detector instability:

1) charging up the surface  by positive ions

and

2) a long-term polarization effect   which may cause quite strong gain variations unless a high voltage is constantly applied to the detector [7,8]. This is why in the next prototype, which is under  manufacture now , the intermediate layer between upper and lower strips will be made from another material having  also a much lower ($\sim 10^{10} \Omega$cm ) resistivity.

Note that attempts to influence the field near the anode of MSGC were done quite a  long time ago, for example some designs of MSGC had a so- called back plane to which the voltage was applied [9]. An MSGC having anode and cathode strips on the opposite sides of the detector substrate was also tested [10]. There were some indications that the inner strips in resistive MICROMEGAS under certain condition also affect the detector gain [11].

All there early studies and our recent observations show that extra electrodes located under or close to the active electrodes can indeed be used  for the tuning  of electric fields in micro pattern detectors. This may offer new possibilities in their design and optimization

## VII. Future perspectives

As a next step in these studies we are planning to implement field correction strips in a multilayer TGEM. A first prototype of such a multi-electrode TGEM with cone holes aiming to better suppress ion back flow [12] has already been built and is under test now. Simulations show (Fig. 13) that by optimizing the voltage across each electrode one can find the optimal field shape that allows electrons  to be brought inside the cones, while blocking ions on the bottom face of the device. An efficient ion back flow blocking could be an attractive feature for several applications: photodetectors, TPC etc.

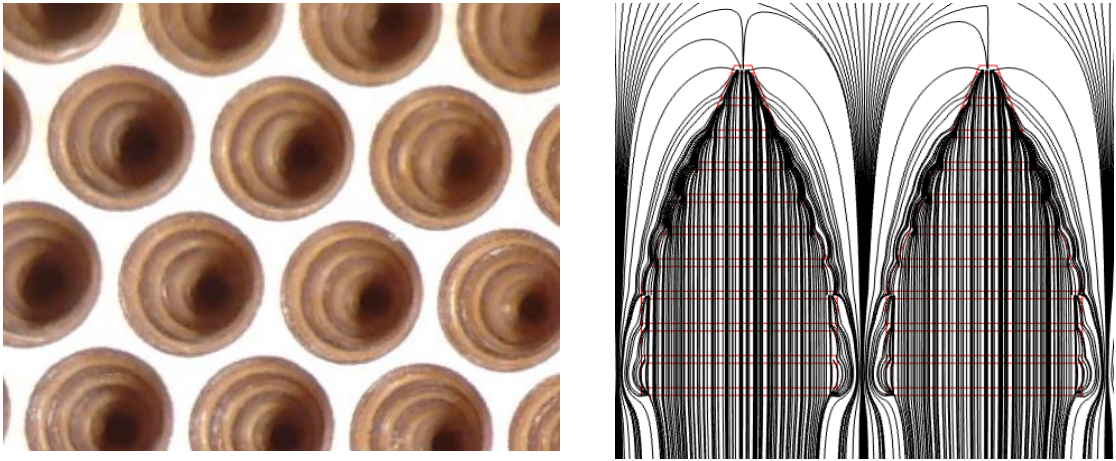

Figure 13. Left: photograph of the first prototype of multielectrode conic TGEM. Right: Simulated fieldlines in this device for a certain voltage configuration on the electrodes. According to this simulation, a large number of ions can be blocked on the bottom face of this device.

**Acknowledgements:**

We would like to warmly thank E. Oliveri, L. Ropelewski for frequent discussions and help throughout this work